\documentclass[10pt,conference,anonymous=true]{IEEEtran}

\usepackage[numbers]{natbib}
\usepackage{graphicx}
\usepackage{placeins}
\usepackage{float}
\usepackage{hyphenat}
\usepackage{xstring}
\usepackage{forloop}
\usepackage{booktabs}
\usepackage{array}
\usepackage{multirow}
\usepackage{multicol}
\usepackage{wasysym}
\newcolumntype{M}[1]{>{\centering\arraybackslash}m{#1}}
\usepackage{tcolorbox}
\tcbuselibrary{breakable}
\usepackage{changepage}
\usepackage{tabularx}
\usepackage{colortbl}
\usepackage{tabularx}
\usepackage{pifont}
\newcommand{\cmark}{\ding{51}}%
\newcommand{\xmark}{\ding{55}}%

\title{SoK: Human-Centered Phishing Susceptibility}
\author{\IEEEauthorblockN{Sijie Zhuo, Robert Biddle, Yun Sing Koh, Danielle Lottridge, Giovanni Russello}\\
\IEEEauthorblockA{\textit{University of Auckland, New Zealand} \\
szhu842@aucklanduni.ac.nz, \{robert.biddle, y.koh, d.lottridge, g.russello\}@auckland.ac.nz}
}

\begin{document}

%


\maketitle
\thispagestyle{plain}
\pagestyle{plain}
\begin{abstract}
Phishing is recognised as a serious threat to organisations and individuals. While there have been significant technical advances in blocking phishing attacks, people remain the last line of defence after phishing emails reach their email client. Most of the existing literature on this subject has focused on the technical aspects related to phishing. However, the factors that cause humans to be susceptible to phishing attacks are still not well-understood. To fill this gap, we reviewed the available literature and we propose a three-stage Phishing Susceptibility Model (PSM) for explaining how humans are involved in phishing detection and prevention, and we systematically investigate the phishing susceptibility variables studied in the literature and taxonomize them using our model. This model reveals several research gaps that need to be addressed to improve users' detection performance. We also propose a  practical impact assessment of the value of studying the phishing susceptibility variables, and quality of evidence criteria. These can serve as guidelines for future research to improve experiment design, result quality, and increase the reliability and generalizability of findings.

\end{abstract}


%
\IEEEpeerreviewmaketitle

\section{Introduction}

Phishing is a form of cyberattack that aims at obtaining the victim's sensitive information and credentials (such as social security number and bank details) or execute malicious code by deceiving the user to perform specific actions such as clicking on the embedded links, downloading, or executing the attached files. Spear-phishing and whaling \cite{alabdan2020phishing} are dangerous because they target a specific group of people and can result in a higher phishing success rate \cite{burda2020don, burns2019spear}. The most common form of phishing is to imitate a legitimate email's visual presentation and content to make users believe that it comes from a trusted source, thus deceiving the target into performing actions that attackers desire (usually clicking an embedded link) \cite{chaudhry2016phishing}. 


There is a growing body of literature on user-centred phishing susceptibility to understand why users fall for phishing attacks. Jampen et al. \cite{jampen2020don} focused on anti-phishing training, and suggested methodologies to design sustainable and effective training to help users minimize their susceptibility to phishing attacks. Sommestad and Karlzén's meta-analysis \cite{sommestad2019meta} investigated  phishing related variables and the quality and flaws that exist in current research. Norris et al. \cite{norris2019psychology} conducted a systematic review on fraud victimization, which included phishing as one of their research areas. Their work focused on  psychological principles to explain how the users' decision-making processes are influenced by the message, as well as experiential and dispositional factors. 

There is still a lack of research work that taxonomizes the phishing susceptibility variables in a systematized way. In order to create such a taxonomy, this paper aims to collect and analyze the key research in relation to phishing susceptibility. We ask the following research questions:
\begin{itemize}
    \item \textit{What variables influence the users' phishing susceptibility?  
    \item What opportunities for support are suggested by these variables, to help reduce susceptibility?
    \item Where are the research gaps?}
\end{itemize}

\begin{figure}[h]
\centering
\includegraphics[width=0.48\textwidth]{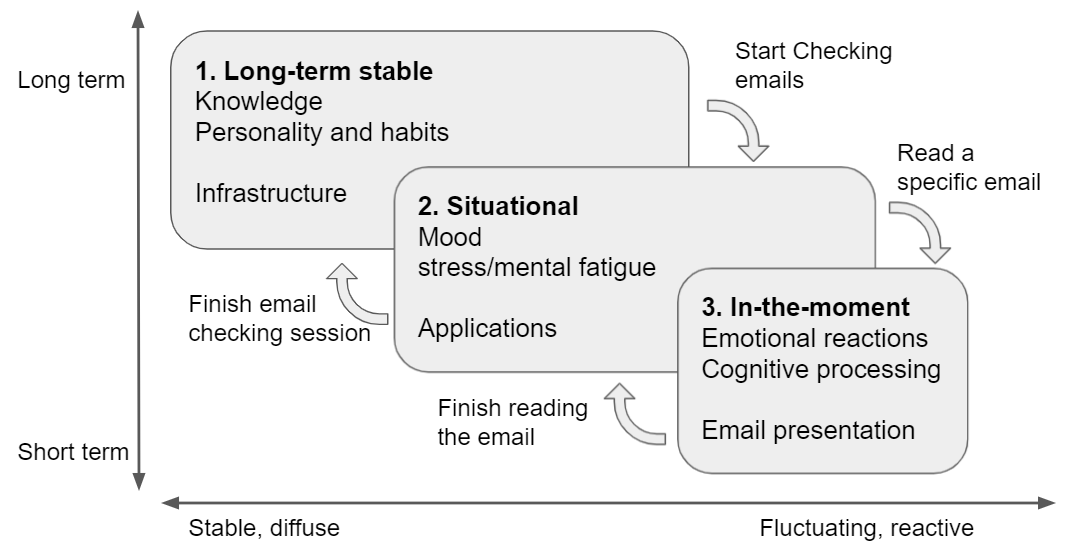}
\caption{The Phishing Susceptibility Model (PSM)}
 \label{fig:model}
\end{figure}

Our research questions motivate us to develop a human-centred model, named \textbf{Phishing Susceptibility Model (PSM)}, that describes the phishing attack process and provides a foundation for analyzing phishing susceptibility variables, as shown in Figure \ref{fig:model}. Our model is a closed-loop. Most of the time, the users would be in stage one, and whenever the users check email, they enter stage two, and once they finish checking emails, they go back to stage one. While in stage two, when the users read a potential phishing email, they enter stage three, and they will return to stage two once they close that email. While checking email, the users can perceive information associated with phishing or misclassify legitimate emails as phishing, which can change their subsequent behaviour. Once the users fall for a phishing attack, there will be two possible consequences: 1) the users do not know that they have been phished, and hence their behaviour does not change; 2) they find out they have been phished, then this experience will update their knowledge, beliefs and attitudes towards phishing (updating stage one), hence impacting future behaviour. Conceptualising phishing in these stages can help us identify opportunities for support at each stage. We believe that our model can provide a better understanding of the human experience in the security protection chain as it covers all stages of where users are associated with phishing. 

The PSM serves as a guide for understanding how the users are involved in phishing detection and prevention. At the same time, it reveals research gaps and provides inspiration and a basis for future human-centred phishing research. Using our model, we found a lack of research on the situational variables that influence users' performance in a particular email checking session. We argue that it is essential to study these situational variables to understand why and under what circumstances the users are more susceptible to phishing attacks. 

In this paper, we bring a particular framing of practical impact, where we discuss the opportunities for solutions related to reducing susceptibility. 

We also contribute a quality of evidence analysis. We rate the quality of evidence based on how confident we can be in the validity and the reliability of findings. This analysis provides guidelines for future research to inform  experiment design quality. By reporting the analysis, our hope is that future studies should be designed knowing which experiment decisions can lead to producing reliable data. To the best of our knowledge, we are the first to propose such an analysis in a systematic review of phishing susceptibility.

In exploring these facets of phishing susceptibility, we make the following contributions:
\begin{itemize}
    \item We introduce the PSM to explain the influence of different variables, and identify research gaps and future research directions using the model.
    \item We taxonomize phishing susceptibility variables that have been studied in the community, and detail the impact of these variables on phishing detection.
    \item We discuss phishing susceptibility variables with regards to the practical opportunities for reducing phishing susceptibility.
    \item We develop a quality of evidence analysis to assess the experiment's quality, and suggest criteria for future research to improve experiment design.\\
\end{itemize}

The rest of this paper is organized as follows. In Section \ref{Methodology}, we explain how we found the relevant literature, and propose a quality of evidence analysis for assessing the quality of the experiment result. In Section \ref{model}, we introduce the PSM we propose, then systematize the phishing susceptibility variables that have been studied in the literature, and categorize them into the PSM. In Section \ref{gaps}, we discuss the research gaps identified using the PSM, and suggest future research directions. Lastly, in Section \ref{conclusion}, we present conclusions and suggest future directions.

\section{Methodology} \label{Methodology}

This section will explain the steps we took to find the 45 phishing relevant papers in the area, and explain the criteria we used to assess each paper we reviewed regarding their radical practical impact and experiment quality.

\subsection{Review paper selection}
We followed a systematic methodology to search for the relevant human-centred phishing research paper. The first step was defining how and where to search for these papers. We selected the following keywords from a list of 11 libraries, which included (ScienceDirect, ACM Digital Library, IEEE Xplore, Scophos, ProQuest, SAGA Journals, Springer Link, Web of Science, Elsevier (INSPEC), CoteSeer, and the AIS elibrary):

\textit{Phishing AND (people OR adults OR human OR employees OR students OR users OR women OR men OR participants OR subjects) AND (experiment* OR study OR studies OR “field trial”) AND [Publication Date: (2000 TO 2020)]}

The search result produced 4,323 papers. We then used an inclusion criteria to filter out the papers that did not meet our requirements. The inclusion criteria ensure that we only included the papers that were: (1) written in English, (2) scientific work, (3) conducting human-centred experiments, (4) involving more than 20 participants, and (5) related to phishing susceptibility. After the filtering process, only 45 papers matched our requirements.

\subsection{Opportunities for impact}
As we analyzed the phishing susceptibility variables, we realized that the study of these variables does not always lead to solutions to help users improve their performance against phishing attacks. For instance, suppose research found that some gender groups or age groups are more susceptible to phishing attacks than others; the next question would be ``so what?". It is not possible to stop these groups of users from using email, and it is not likely that companies would refuse to hire these groups of people just because of their higher susceptibility to phishing attacks. Therefore, we see a need to discuss each phishing susceptibility variable with an impact score to assess how changeable these variables are in helping users reduce their phishing susceptibility. Further, this rating should be used for prioritizing the area of human-centred phishing studies. Research should focus more on the variables that can be changed, such as the amount of knowledge the users have, instead of the variables users do not control, such as their personalities and gender. In this paper, we will be rating  each variable with \textit{few, medium or many} opportunities for impact.

\begin{tcolorbox}[breakable]
\textbf{Take Away 1:} Future research should consider practicality of impact when exploring research areas because the studies that lead to actionable changes are more valuable in protecting users from phishing attacks.
\end{tcolorbox}

\subsection{Quality of evidence} \label{quality}

For phishing experiments, reliable and generalizable data is essential for deriving high quality and convincing results, and thus the confidence of the findings. By studying the experiment design of the reviewed literature, we found a large variation in the quality of the experiments, which could partially explain the inconsistencies that exist in research findings. To address this issue, we propose a quality of evidence criteria for assessing the experiment design quality. As demonstrated in Table \ref{tab:factorCoverage}, we illustrate the use of these criteria with the existing literature and provide an overview of the experiment quality in the literature. These criteria also serve as a guideline for future research to consider how the experiment design choices should be made to collect sufficient and high-quality data. Our quality of evidence analysis consists of two dimensions: \textit{experiment type and sample size}, which we explain below. Please note that we do not mean to disparage the potential insight developed from these studies: our comments only affects the confidence of using the findings in future work in the related area.


\subsubsection{Experiment type}


For researchers, there are mainly two ways of conducting human-centred phishing experiments, via a phishing simulation study and email management study. The phishing simulation approach is also known as embedded training, and has been popularly used in industry. This approach is usually carried out in the real-world environment. It can closely capture users' real-world behaviour and provides realistic and reliable feedback on the quality of the ``attack" and the users' performance.  Note that the phishing emails' quality, relevance and presentation directly affect the click rate and phishing success rate of the study. This type of experiment has high ecological validity. The drawback of this kind of experiment lies in the amount of data that can be acquired per participant. Since not all participants will ``fall for the attack", extra participants would be needed to produce sufficient ``victims" for further analysis. Also, research involves complex ethical and legal issues because conducting a realistic phishing campaign may involve imitating emails from third-party identities.

An email management study involves participants managing a list of predefined emails. For each email the participants manage, they will be asked to either select how they want to respond to the email (multi-choice questions) \cite{sheng2010falls}, or judge the email's legitimacy \cite{lawson2020email}. Email management studies excel at collecting a large quantity of data regarding participants' responses to different emails. The number of emails directly contributes to the amount of data can be collected. Based on the literature reviewed, the average number of emails chosen in a study is around 40. Email management studies are usually conducted in a controlled environment for precise control of the variables, and to explore stronger relationships between the tested variables. However, the nature of the experiment suggests that participants' decisions may not reflect real-world behaviour. They may be more cautious because they are participating in a study. Alternatively, because there are no real-world consequences related to their decisions, they may also be less cautious \cite{downs2007behavioral}. Furthermore, these experiments usually have limitations in presenting the email. Since participants may have different preferences or constraints on the email client and environment setup, the experiment material may not reflect the participants' real working environment, which could influence the result's reliability.

Apart from these two types of experiments, studies may also collect data through other research methods such as surveys. Self-selected and self-reported surveys and questionnaires are the most cost-efficient method of collecting user data. However, these measures may not accurately reflect the real-world population or behaviour due to response bias. In the reviewed papers, 6 studies use surveys as their only method for measuring the participants' phishing susceptibility, and based their findings on the self-reported data. Hence, we consider survey-only study as a third methodology for studying phishing susceptibility.


\subsubsection{Sample size}
The sample size is another essential measure that contributes to the statistical significance of the study findings. As suggested by Sommestad and Harlzen \cite{sommestad2019meta}, power analysis \cite{cohen2013statistical} is necessary to determine the minimal sample size required to produce a significant result. Even though studies may want to recruit as many participants as possible, we suggest that power analysis should be performed when possible as a sanity check. Further, we consider measuring the effect size of the sample population as a better option because statistical significance only tests the existence of statistical difference between populations, whereas effect size more focus on the magnitude of the difference \cite{sullivan2012using}. In other words, even if a statistical significance is found between sample groups, if the effect size is small, the result may not lead to meaningful findings. Therefore, we see a need for future research to carry out effect size analysis even with large sample population studies to strengthen the argument. However, in the last twenty years, effect size is almost never reported in the literature, hence we can only use the sample size as a measure for assessing the quality of the findings.

\begin{figure}[h!]
\centering
\includegraphics[width=0.48\textwidth]{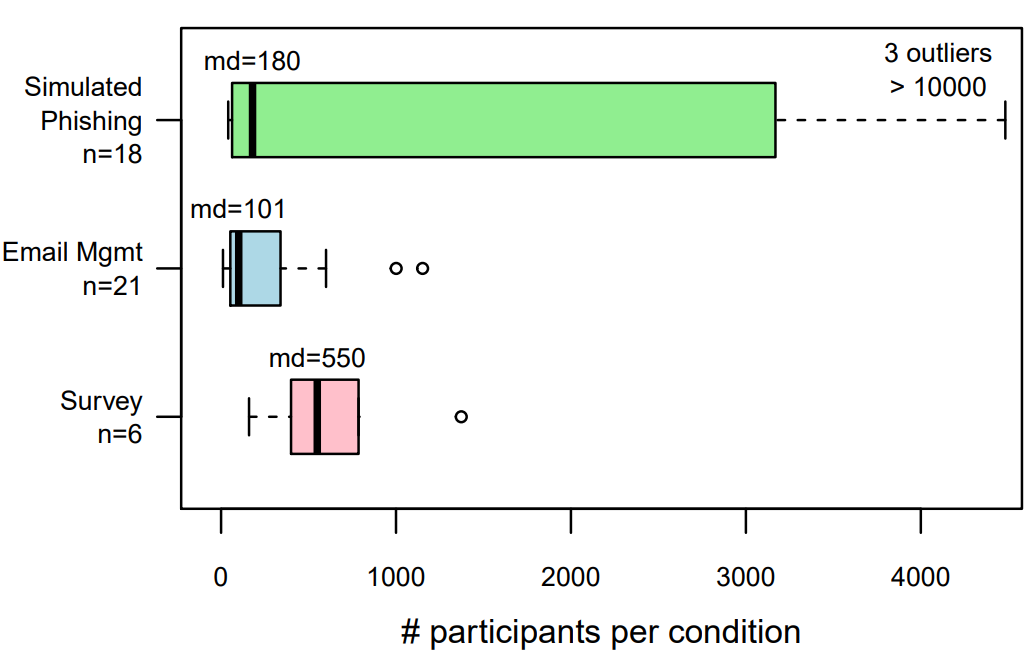}
\caption{Distribution of experiment sample size by type of study}
 \label{fig:samplesize}
\end{figure}

Due to the different experiment designs across the studies, it is infeasible and unreasonable to provide a sample size standard for all studies. For example, since the click rate is unknown for simulated phishing attacks, more participants would be needed to ensure the ones that fall for the attack are numerous enough for further analysis. Hence, in this paper, as demonstrated in Table \ref{tab:samplesize}, we present a relative rating base on existing literature to assess their group sample size (the average sample size per condition group). The studies are first categorized by their experiment design, where their sample size criteria are rated separately. For each category, about 30\% of the reviewed studies with the largest group sample size are classified as having a large (\textit{L}) sample size, followed by about 40\% as medium (\textit{M}) and the remaining 30\% are classified as having a small (\textit{S}) sample size. 
Figure \ref{fig:samplesize} shows distributions of sample sizes of the main studies in the papers reviewed. The boxplots show the median (dark vertical line), inner quartiles (coloured box), outer quartiles (whiskers), and outliers (circles).

\begin{table}[h!]
\centering
\caption{Group sample sizes for associated research methods \label{tab:samplesize}}
\begin{tabular}{c|c|c|c} 
\toprule
\multirow{2}{*}{Experiment type} & \multicolumn{3}{c}{Group sample size}  \\ 
\cline{2-4}
                                 & L & M & S                  \\ 
\hline
Simulated phishing experiment    & $> 1000$  & 1000 - 70  & $< 70$                  \\
Email management study      & $> 300$  & 300 - 55  & $< 55$                    \\
Survey only study         & $> 1000$  & 1000 - 350  & $< 350$                    \\
\bottomrule
\end{tabular}
\end{table}

\section{Results} \label{model}

To systematize the variables that relate to influencing users' phishing susceptibility, we review findings reported in the literature and summarize them in Table \ref{tab:factorCoverage}. As we reviewed the literature, we found three key stages for variables relating to phishing email attacks: (1) stable characteristics such as age, personality, which form a baseline before the users check their emails; (2) variables that may change depending on the situation, such as whether they check their email on a desktop or a smartphone; and (3) in the moment characteristics when they read a potential phishing email.  Accordingly, in Table \ref{tab:factorCoverage}, we group the literature based on these temporal scales: Stage one as the long-term stable state; Stage two as the situational state; and Stage three as the in-the-moment state. Additionally, for each study we reviewed, we present a subjective rating of how centred the research was on the variable in question. The glyph $\CIRCLE$ indicates that the variable was investigated as a primary research goal, $\LEFTcircle$ for a secondary research goal, and $\Circle$ for only presenting the result. Articles may have included one or more variables, in primary or secondary capacities, in a single experiment or across several experiments. When looking at Table \ref{tab:factorCoverage}, certain patterns emerge. The Stage one variables are relatively easy to determine, and we see Stage one variables are prevalent in the literature. We see a gap in Stage two variables, with relatively few existing studies on situational variables that may influence phishing susceptibility. We observe that Stage three have the most full circles, which means that there has been more focus on cognitive effort, persuasive methods, and visual presentation in the phishing literature. The rows of the table are ordered to group studies on Stage one variables first, then Stage two, and then Stage three. We also ordered the table by date of study publication, but did not observe any patterns. The last column in Table \ref{tab:factorCoverage} presents the quality of evidence using two criteria: experiment type and sample size. The entries in the column represent the group sample size for type of experiment, as summarized in Table \ref{tab:samplesize}. \footnote{Note, the sample size for each experiment types are different, a phishing simulation study rated as small could involve more participants than a email management study rated with medium sample size}

\begin{table*}[thbp]
\centering
\caption{Systemization of current literature and coverage of factors among the studies}
\label{tab:factorCoverage}
\begin{tabular}{M{4cm}|M{0.7cm}M{0.7cm}M{0.7cm}|M{0.7cm}M{0.7cm}|M{0.7cm}M{0.7cm}M{0.7cm}|M{0.7cm}|M{0.7cm}|M{0.7cm}} 
\toprule
\multirow{2}{*}{\textbf{Literature}} & \multicolumn{3}{c|}{\textbf{Long term stable}}                               & \multicolumn{2}{c|}{\textbf{Situational}}                & \multicolumn{3}{c|}{\textbf{In the moment}}                                   & \multicolumn{3}{c}{\textbf{Quality (sample size)}} \\ 
\cline{2-12}
                       & \rotatebox{90}{\parbox{1.75cm}{Knowledge}} & \rotatebox{90}{\parbox{1.75cm}{Demographics}} & \rotatebox{90}{\parbox{1.75cm}{Personality\\and habits}} & \rotatebox{90}{\parbox{1.75cm}{Access\\method}} & \rotatebox{90}{\parbox{1.75cm}{Situational\\characteristics}} & \rotatebox{90}{\parbox{1.75cm}{Cognitive\\effort}} & \rotatebox{90}{\parbox{1.75cm}{Persuasion\\methods}} & \rotatebox{90}{\parbox{1.75cm}{Visual\\presentation}} & \rotatebox{90}{\parbox{1.75cm}{Phishing\\simultion}} & \rotatebox{90}{\parbox{1.75cm}{Email\\management}} & \rotatebox{90}{\parbox{1.75cm}{Survey\\only}}\\ 
\hline
Musuva et al. \cite{musuva2019new}                       &\CIRCLE&\Circle&\Circle&   &\Circle&\CIRCLE&\LEFTcircle&\LEFTcircle&L  &  &   \\\arrayrulecolor{gray}\hline
Parsons et al. \cite{parsons2013phishing}                &\LEFTcircle&\Circle&\LEFTcircle&   &   &\CIRCLE&   &   &   &M  & \\\arrayrulecolor{gray}\hline
Wang et al. \cite{wang2017coping}                        &\CIRCLE&\Circle&\Circle&   &   &\CIRCLE&   &   &   &L  & \\\arrayrulecolor{gray}\hline
Janet et al. \cite{janet2008analysis}                    &\CIRCLE&\Circle&\Circle&   &   &   &   &\Circle&   &   &S \\\arrayrulecolor{gray}\hline
Tjostheim and Waterworth \cite{tjostheim2020predicting}  &\Circle&\Circle&\CIRCLE&   &   &   &   &   &   &   &L \\\arrayrulecolor{gray}\hline
Burda et al. \cite{burda2020testing}                     &\Circle&\CIRCLE&   &   &\Circle&\Circle&\CIRCLE&\LEFTcircle&M  &   & \\\arrayrulecolor{gray}\hline
Jalali et al. \cite{jalali2020employees}                 &\CIRCLE&\Circle&   &   &\LEFTcircle&   &   &   &L  &   & \\\arrayrulecolor{gray}\hline
Wang et al. \cite{wang2016overconfidence}                &\CIRCLE&\Circle&   &   &   &\Circle&\Circle&   &   &L  & \\\arrayrulecolor{gray}\hline
Alseadoon et al. \cite{alseadoon2013typology}            &\Circle&\CIRCLE&   &   &   &\Circle&   &   &M  &   & \\\arrayrulecolor{gray}\hline
Diaz et al. \cite{diaz2020phishing}                      &\CIRCLE&\Circle&   &   &   &\CIRCLE&   &   &M  &   & \\\arrayrulecolor{gray}\hline
Mohebzada et al. \cite{mohebzada2012phishing}            &\Circle&\CIRCLE&   &   &   &\Circle&   &   &L  &   & \\\arrayrulecolor{gray}\hline
House and Raja \cite{house2019phishing}                  &\CIRCLE&\Circle&   &   &   &   &\CIRCLE&       &S  &   & \\\arrayrulecolor{gray}\hline
Lin et al. \cite{lin2019susceptibility}                  &\LEFTcircle&\LEFTcircle&   &   &   &   &\CIRCLE&   &M  &   &  \\\arrayrulecolor{gray}\hline
Taib et al. \cite{taib2019social}                        &\LEFTcircle&\LEFTcircle&   &   &   &   &\CIRCLE&   &L  &   & \\\arrayrulecolor{gray}\hline
Baillon et al. \cite{baillon2019informing}               &\CIRCLE&\LEFTcircle&   &   &   &   &   &   &L  &   & \\\arrayrulecolor{gray}\hline
Manasrah et al. \cite{manasrah2015toward}                &\CIRCLE&\LEFTcircle&   &   &   &   &   &   &   &   &M \\\arrayrulecolor{gray}\hline
Sarno et al. \cite{sarno2017phishers}                    &\LEFTcircle&\CIRCLE&   &   &   &   &   &   &   &M  & \\\arrayrulecolor{gray}\hline
Sheng et al. \cite{sheng2010falls}                       &\CIRCLE&\CIRCLE&   &   &   &   &   &   &   &L  & \\\arrayrulecolor{gray}\hline
Vishwanath et al. \cite{vishwanath2011people}            &\LEFTcircle&   &\LEFTcircle&   &\LEFTcircle&\CIRCLE&\LEFTcircle&\LEFTcircle&    &   &S \\\arrayrulecolor{gray}\hline
Petelka et al. \cite{petelka2019put}                     &\Circle&   &\Circle&   &\LEFTcircle&\CIRCLE&   &\CIRCLE&   &M  & \\\arrayrulecolor{gray}\hline
Sarno and Neider \cite{sarno2021so}                      &\LEFTcircle&   &\LEFTcircle&   &\CIRCLE&   &   &   &   &S  & \\\arrayrulecolor{gray}\hline
Pfeffel et al. \cite{pfeffel2019user}                    &\Circle&   &\Circle&   &   &\Circle&   &\CIRCLE&   &S  & \\\arrayrulecolor{gray}\hline
Vishwanath et al. \cite{vishwanath2018suspicion}         &\CIRCLE&   &\LEFTcircle&   &   &\CIRCLE&   &   &M  &   & \\\arrayrulecolor{gray}\hline
Williams and Polage \cite{williams2019persuasive}        &\Circle&   &\Circle&   &   &   &\CIRCLE&\CIRCLE&   &S  & \\\arrayrulecolor{gray}\hline
Alseadoon et al. \cite{alseadoon2015influence}           &\Circle&   &\CIRCLE&   &   &   &   &\CIRCLE&M  &   & \\\arrayrulecolor{gray}\hline
Welk et al. \cite{welk2015will}                          &\LEFTcircle&   &\CIRCLE&   &   &   &   &   &   &S  & \\\arrayrulecolor{gray}\hline
Gordon et al. \cite{gordon2019assessment}                &\LEFTcircle&   &   &   &\Circle&   &\CIRCLE&   &L  &   & \\\arrayrulecolor{gray}\hline
Harrison et al. \cite{harrison2016individual}            &\Circle&   &   &   &   &\CIRCLE&\CIRCLE&\Circle&S  &   & \\\arrayrulecolor{gray}\hline
Canfield et al. \cite{canfield2016quantifying}           &\CIRCLE&   &   &   &   &\CIRCLE&   &   &   &S  & \\\arrayrulecolor{gray}\hline
Jansen and Van \cite{jansen2018persuading}               &\LEFTcircle&   &   &   &   &\CIRCLE&   &   &   &   &M \\\arrayrulecolor{gray}\hline
Molinaro and Bolton \cite{molinaro2019using}             &\Circle&   &   &   &   &\CIRCLE&   &   &   &M  & \\\arrayrulecolor{gray}\hline
Parsons et al. \cite{parsons2016users}                   &\Circle&   &   &   &   &   &\CIRCLE&\CIRCLE&   &M  & \\\arrayrulecolor{gray}\hline
Downs et al. \cite{downs2007behavioral}                  &\CIRCLE&   &   &   &   &   &   &\Circle&   &M  & \\\arrayrulecolor{gray}\hline
Canfield et al. \cite{canfield2019better}                &\CIRCLE&   &   &   &   &   &   &   &   &L  & \\\arrayrulecolor{gray}\hline
Perrault \cite{perrault2018using}                        &\CIRCLE&   &   &   &   &   &   &   &   &L  & \\\arrayrulecolor{gray}\hline
Chuchuen and Chanvarasuth \cite{chuchuen2015relationship}&   &\Circle&\CIRCLE&   &   &   &   &\LEFTcircle&   &   &M \\\arrayrulecolor{gray}\hline
Blythe et al. \cite{blythe2011f}                         &   &\Circle&   &\LEFTcircle&   &   &\CIRCLE&\LEFTcircle&   &M  & \\\arrayrulecolor{gray}\hline
Sarno et al. \cite{sarno2020phish}                       &   &\CIRCLE&   &   &\Circle&\LEFTcircle&   &   &   &S  & \\\arrayrulecolor{gray}\hline
Wright et al. \cite{wright2014research}                  &   &\Circle&   &   &   &\Circle&\CIRCLE&   &S  &   & \\\arrayrulecolor{gray}\hline
Goel et al. \cite{goel2017got}                           &   &\LEFTcircle&   &   &   &   &\CIRCLE&   &M  &   & \\\arrayrulecolor{gray}\hline
Lawson et al. \cite{lawson2020email}                     &   &   &\CIRCLE&   &   &   &\CIRCLE&   &   &M  & \\\arrayrulecolor{gray}\hline
Curtis et al. \cite{curtis2018phishing}                  &   &   &\CIRCLE&   &   &   &   &   &   &L  & \\\arrayrulecolor{gray}\hline
Harrison et al. \cite{harrison2015examining}             &   &   &   &   &   &\CIRCLE&   &\CIRCLE&S  &   & \\\arrayrulecolor{gray}\hline
Tian and Jensen \cite{tian2019effects}                   &   &   &   &   &   &   &\CIRCLE&   &S  &   & \\\arrayrulecolor{gray}\hline
Arduin \cite{arduin2020click}                            &   &   &   &   &   &   &   &\CIRCLE&   &M  & \\

\bottomrule
\end{tabular}
\end{table*}

We created the Phishing Susceptibility Model (PSM) (Figure \ref{fig:model}) to summarise the variables that are relevant to phishing susceptibility in Table \ref{tab:factorCoverage}. 
When we refer to users, we mean both employees in organisations or individuals on personal devices, where typically both cases are relevant. Phishing email typically attempts to influence user behaviour beyond the email itself, for example by interacting with a website or opening an attachment. Phishing is both relevant to organizations because they are protecting their organizational resources and to individuals because they are protecting their properties and privacy. We consider interactions beyond the phishing email outside our scope and do not directly address follow-up actions in our model. The issues that arise become more diverse, relating to topics such as browser indicators or operating system administration. We acknowledge these topic issues are important, but our focus is only on the susceptibility of users to engage beyond the phishing email itself.






\subsection{Stage one: long-term stable }

Stage one consists of the users' long-term stable variables that shape their basic responses towards phishing attacks; it also refers to the individual differences among the users. For instance, individuals with higher impulsivity may respond to emails quickly without paying much attention \cite{lawson2020email, parsons2013phishing}, whereas individuals with more phishing knowledge can perform better in identifying phishing emails \cite{sheng2010falls, harrison2016individual}. It is reasonable to believe that some users are better at detecting phishing emails than others and are less susceptible to phishing. The existing research has mainly focused on the following variables: phishing related knowledge, demographics, personality and habits. As illustrated in Table \ref{tab:factorCoverage}, Stage one factors are the most studied areas. It has been largely agreed that higher phishing related knowledge can reduce phishing susceptibility, hence better phishing detection performance. 

\subsubsection{Knowledge}

From Table \ref{tab:factorCoverage}, knowledge is the most analysed phishing susceptibility variable. Knowledge provides the foundation for phishing detection. Without knowledge, users cannot distinguish between legitimate and phishing emails. Sheng et al.'s study \cite{sheng2010falls} found that participants with more knowledge and more experience from training are less susceptible to phishing. Similar work regarding the effect of knowledge on phishing susceptibility has been discussed in many other studies \cite{alseadoon2015influence, baillon2019informing, downs2007behavioral, wang2017coping, musuva2019new, sarno2017phishers, harrison2016individual, pfeffel2019user, gordon2019assessment, datar2014awareness}, and all confirmed that higher phishing related knowledge leads to higher phishing detection performance. Knowledge of other domains can also help users distinguish between legitimate and phishing emails \cite{wash2020experts}. For example, the individuals who work for a bank should be more familiar with emails from banks, thus, are more likely to pick up the unusual cues in the bank-based phishing emails.

Knowledge can be categorized into explicit and implicit knowledge. Explicit knowledge is usually gained through learning and direct training, whereas implicit knowledge is gained from experiences, especially after encountering phishing attacks. Studies have found that, the more individuals are familiar with computers and technology, the more capable individuals would be in coping with phishing emails \cite{pattinson2012some}. Besides, users with more experience related to information technology and cybersecurity tend to spend more time and effort in checking emails \cite{harrison2016individual, redmiles2018examining}. Baillon et al.'s experiment \cite{baillon2019informing} compared the effectiveness of direct training and embedded training using a simulated phishing campaign. Their results suggest that both types of training can improve users' phishing detection performance, but embedded training is more effective in that the experience of falling for a phishing attack can raise their awareness about phishing for subsequent email checking attempts. This work indicates a benefit from training, however industry reports indicate that training is not effective enough to solve the problem. A report released by Cloudian found that 65\% of organisations that fell victim to phishing attacks actually had trained their staff \cite{cloudian2020}. As pointed out by Jampen et al. \cite{jampen2020don}, continuous training is needed to keep users alert and maintaining high performance in phishing detection. Further, training material needs to be personalized so that users with different knowledge levels can be targeted with different materials to achieve better effectiveness.


We consider knowledge has many opportunities for impact because consistent results have been found in studies that knowledge directly influences phishing susceptibility. We argue that knowledge not only determines how users respond to phishing emails, but this variable can also influence the in-the-moment state, and change users' behaviour. Furthermore, we perceive a potential future direction in studying approaches to help users efficiently gain phishing related knowledge.

\begin{tcolorbox}[breakable]
\textbf{Take Away 2:} Current anti-phishing training is not effective in protecting users from phishing attacks. It is essential for future research to investigate approaches that \textit{truly} help the users reduce phishing susceptibility.
\end{tcolorbox}


As part of phishing knowledge, perception and beliefs and shape our feelings toward phishing attacks: how we perceive threats, how we perceive our efficacy, and our confidence can all influence phishing detection performance. However, most of the findings on these topics are not strong due to the lack of realistic context \cite{canfield2019better, wang2017coping, jansen2018persuading, wang2016overconfidence}, or insufficient amount of participants \cite{house2019phishing}. Nevertheless, the influence of perception and beliefs on users' phishing susceptibility is likely, as these beliefs would act as a booster or suppressor for the motivation of actively engaging with email reading activity. 


Witte et al. \cite{witte1992putting} defined perceived threat as the subjective evaluation of the threat present in a situation. Perceived threat has two components: perceived severity and perceived susceptibility. Perceived severity (also called perceived consequence) is described as one's belief about the magnitude and significance of the threat and the consequence of falling for the threat. Canfield et al.'s email management study \cite{canfield2019better} found that more negative consequences lead to shifting their judgment towards treating more emails as phishing emails. Consequently, even though the belief can reduce the chance of falling for phishing attacks, more false-positive judgements will be made. Also, Wang et al.'s study \cite{wang2017coping} suggests that perceived threat is positively related to phishing anxiety (concern of falling for a phishing attack), such that it would induce high anxiety, leading to risky behaviour. The other component is perceived susceptibility, which is one's belief about how likely the person would fall for a phishing attack. Wang et al.'s study found that the belief of a high likelihood of having been phished can lead to lower detection performance. Interestingly, this contrasts with Vishwanath et al.'s study \cite{vishwanath2018suspicion}, which suggests that these beliefs would alert the users to motivate more systematic processing.

Apart from perceived threat, the individuals' reflection on their own ability to deal with phishing emails also contributes to their motivation to perform recommended responses. Perceived efficacy involves beliefs about the recommended response's effectiveness and how feasible it is for the individual to carry out the recommended action. Correspondingly, perceived efficacy has two dimensions: (1) the response efficacy for describing the individual's beliefs about the effectiveness of the recommended response in dealing with the threat; and (2) the perceived self-efficacy for individuals' beliefs about their ability to carry out the response \cite{witte1992putting}. For response efficacy, as addressed in Jansen et al.'s study \cite{jansen2018persuading}, it is associated with protective motivation. Concerning perceived self-efficacy, higher perceived self-efficacy would lead to higher motivation in performing protective actions against phishing attacks, such as not responding to the emails \cite{house2019phishing, jansen2018persuading}. 

Perceived self-efficacy is a type of confidence belief; it is also referred to as individuals' prospective confidence \cite{busey2000accounts}. Canfield et al. \cite{canfield2016quantifying} found that confidence is strongly related to the individuals' judgement about emails: higher confidence usually leads to a higher tendency to classify an email as legitimate. However, studies also have found that users are usually overconfident in their ability to detect phishing emails than is really justified \cite{wang2016overconfidence, canfield2019better}. Overconfidence can lead to paying less attention to the email, thus exposing them to more danger \cite{wang2016overconfidence}.

Different beliefs can affect user behaviour in different ways. Some beliefs may result in users being too unconcerned about phishing, and thus users may fall for simple attacks. Other beliefs may result in users being suspicious of almost all emails, thus causing many false alarms. We rate this variable with medium opportunities for impact as even though the effect of some beliefs is still unclear, studies have shown that these beliefs do have an impact on phishing susceptibility.  Future research could focus on finding the ``sweet spot" for balancing such beliefs to motivate protective behaviour while still allowing normal email to be processed.

\subsubsection{Demographics} 

The most studied demographic variables are gender and age \cite{baillon2019informing,sheng2010falls, taib2019social, lin2019susceptibility, mohebzada2012phishing, sarno2017phishers, sarno2020phish}. Existing literature has found inconsistent and insignificant results regarding different age and gender groups. Sheng et al.'s study \cite{sheng2010falls} demonstrated an interesting finding. The participants were asked to perform two email management tasks, with a training session in between. Their result showed that for the first email management task, women fell for significantly more phishing emails than men, but the research also points to a confound: that these women had less technical knowledge than the men. After a training session, both men and women performed equally well. The result is evidence that gender does not itself cause a difference in the person's ability in detecting phishing emails, but knowledge of individuals can.

Sarno et al. \cite{sarno2020phish} conducted an email management study that focuses explicitly on phishing susceptibility between younger and older people. This study found no significant difference in detection accuracy between younger and older adults. However, it is interesting that the two age groups used different strategies when managing emails: younger adults were more likely to classify an email as legitimate, whereas older adults were more likely to classify an email as phishing. Further, Taib et al.'s study \cite{taib2019social}, and Lin et al.'s study \cite{lin2019susceptibility} suggest that users of different age groups may be susceptible to different types of phishing. Hence, the relationship between age and phishing susceptibility is still uncertain.


Another demographic issue that has been collected frequently is user occupation. For students, this includes their academic major area of study. Studies \cite{burda2020testing, mohebzada2012phishing, diaz2020phishing, manasrah2015toward, goel2017got} agreed that students or junior employees were more prone to phishing attacks than staff or senior employees. Also, technical knowledge and experience gained from academia or industry can help users reduce phishing susceptibility. Students majoring in IT or engineering (STEM), and professionals in the industry, are less susceptible to attacks \cite{diaz2020phishing, taib2019social, manasrah2015toward}. These findings demonstrate how user occupation can influence the accessibility of phishing related knowledge, thus affect the users' phishing susceptibility. 

From these findings, it appears that gender and age do not \textit{directly} contribute to phishing susceptibility. Still, they could contribute to the development of individuals' cognition and behaviour. Different demographic groups could be associated with different access to knowledge related to phishing. Further, these demographic variables are difficult to change, and users may not have control of these changes. However, we see an opportunity to study targeted training for specific demographic groups to help reduce their phishing susceptibility. Therefore, we consider demographic variables as having medium opportunities for impact.

\begin{tcolorbox}[breakable]
\textbf{Take Away 3:} We conclude that demographics is an indirect factor of phishing susceptibility, but it can help capture the trends of knowledge and suggest development of targeted training for specific populations in need.
\end{tcolorbox}

\subsubsection{Personality and habits}

Apart from knowledge and demographics, personalities and habits are also considered as long-term stable variables. Studies show that these variables can influence phishing susceptibility.

Personality has been a popular area of research. Many studies \cite{lawson2020email, welk2015will, alseadoon2015influence, pattinson2012some} have assessed the phishing victims' personalities using the well-established Big Five personality traits \cite{john1999big} (\textit{extroversion, agreeableness, openness, conscientiousness, and neuroticism}) to investigate which traits lead to more susceptibility. Yet, these studies found contradictory results. Apart from the Big Five personality traits model, other personality models such as ``the dark triad" (\textit{Machiavellianism, narcissism, and psychopathy}) \cite{curtis2018phishing}, or the DISC model (\textit{dominance, influence, steadiness and conscientiousness}) \cite{chuchuen2015relationship} have been adopted by researchers for a similar purpose. However, the research findings of these two studies are relatively weak compared with the studies that adopted the Big Five traits, because there is a lack of studies using the same metrics, the studies are conducted with no realistic context, and only involve small groups of participants. 


Individuals' ability to regulate their emotions can also contribute to their performance in detecting phishing emails. 
For instance, high impulsivity can lead to poorly conceived, and risky behaviours \cite{evenden1999varieties}. Several studies \cite{lawson2020email, welk2015will, parsons2013phishing, tjostheim2020predicting} have used the Cognitive Reflection Test (CRT) in their phishing study to assess individuals' impulsivity. These studies found that individuals with lower impulsivity (therefore good impulse regulation) performed better in detecting phishing emails. 

Overall, more than half of these studies suffer from either low sample size or unrealistic context. We consider personality to have medium opportunities for impact because even though many of the personality studies found insignificant or inconsistent findings, this area of research can also help identify potential population groups that are more susceptible to phishing, so that more targeted training can be developed and carried out to help these users. However, there is a need for more high-quality studies to validate the existing findings.

Several studies have focused on email reading habits. As suggested by the ``principle of least effort" \cite{zipf2016human}, people tend to use the most convenient and least effortful mode when making decisions. Email reading habits can be formed by frequent access to emails to build up a routine workflow to reduce the cognitive effort required to respond to the emails correctly \cite{vishwanath2011people}. Williams et al.'s study \cite{williams2019persuasive} shows that habituated email usage can lead to a higher tendency of responding instead of ignoring email. Though, it is worth noting that not all habituated reading processes lead to higher phishing susceptibility. Wash's interview \cite{wash2020experts} with security experts found that there was one particular expert that always hovered over every embedded link in emails to check legitimacy. Therefore, we perceive the study of email reading habits as having medium opportunities for impact because it would be valuable to investigate approaches to help users build up good habits to reduce phishing susceptibility.\\

Among the Stage one variables, despite the differences in design quality, it is certain that more phishing related knowledge leads to lower phishing susceptibility, and it is an actionable predictor of phishing susceptibility with good opportunities for impact. The studies of other variables, such as demographics and personalities, found insignificant and inconsistent results. Further, these are stable characteristics with fewer opportunities for impact. Therefore, future research should focus on helping users gain phishing related knowledge efficiently and effectively, and carry out frequent user training to help them maintain a high level of awareness and detection performance.

\subsection{Stage two: situational state}

As users may check their emails under different situations and environments, their long-term stable characteristics may suggest various behavioural patterns. For example, one study \cite{hygge2001effects} has shown that workers tend to work faster in a noisy environment, but with a cost of reduced quality. In the context of phishing, spending less time on each email could lead to decreasing the phishing detection accuracy. Stage two considers the variables that are situated to a particular email checking session. For instance, how users perceive email information, and how their surroundings influence their judgements, are considered in this stage. As shown in Table \ref{tab:factorCoverage}, Stage two factors are the least studied factors. More research in this domain might uncover some interesting insights.  

\subsubsection{Access method}
How individuals access their emails can account for the information acquired from the email, thus influence phishing susceptibility. One interesting study \cite{blythe2011f} investigated how blind people respond to phishing attacks. The result shows that blind people are significantly better at identifying phishing emails than other users. Due to their visual impairment, blind people have to rely on their screen readers to read out the message. This audio representation of the information can minimize the effect of visual distractions (from multimedia) and allow the user to only focus on the main message. The audio representation makes spelling mistakes and suspicious cues more prominent, thus easing phishing detection. The study highlighted how the interpretation of messages could impact the information users perceive, and hence affect their decision making. We consider this variable as having many opportunities for impact because there are many ways information could be presented to users, and some choices may make detection easier. (Also see our discussion of visual presentation in Stage three, below.)

\subsubsection{Situational characteristics}
The situational characteristics refer to the external stimuli/variables that could influence phishing susceptibility, such as email load, workplace management, and email distribution time. These variables might not be controllable for individual users, but some of them might still be changed to help users against phishing.

Email load describes the volume of email the users received in a time period. It is hypothesized that with a high email load, the users would pay less attention to each email, hence reduce phishing detection performance \cite{vishwanath2011people}. Vishwanath et al.'s experiment \cite{vishwanath2011people} does support his hypothesis, in that participants experiencing a high volume of emails are significantly more likely to respond to phishing emails. Sarno et al.'s study \cite{sarno2021so} also found a supportive result that high email load may negatively influence how users classify emails. Similarly, Jalali et al.'s experiment \cite{jalali2020employees} found that high workload is associated with lowering the tendency to follow security policies, hence leading to higher risk. However, Musuva et al.'s study \cite{musuva2019new} found the contradictory result that participants under a higher volume of emails were less susceptible to phishing. Future research could investigate what causes the different behaviour under similar situations. Since the users have no control over how many emails arrive in their inboxes, we label this variable as having medium opportunities for impact. It might be worth exploring design solutions that structure email from different sources distinctly, which might influence phishing susceptibility.

One study targeting organizations \cite{jalali2020employees} found that, workplace management can influence employees' motivation to perform protective behaviours that could help reduce phishing susceptibility. By understanding the purpose of their companies' information security policies and trusting in their management, employees are more willing to accept that following those policies will help them protect the company. As a result, this positive atmosphere means they may be more inclined to follow the policies and carry out protective practices. We rate this variable as having medium opportunities for impact as this approach is actionable, but it requires changes at the management team level to influence the employees' behaviour.

Phishing email's distribution time also contributes to phishing susceptibility. Phishing attacks are not uniformly distributed across the year. As been found in Oest et al.'s study \cite{oest2020sunrise}, phishing attacks occur more frequently near holidays. Gordon et al. \cite{gordon2019assessment} found that the link click-through rate was lower during spring and summer and higher during autumn. The click rate difference may be due to the different amount of phishing emails received at a particular period throughout the year. For the users, it would be difficult to know when the next phishing email might come. Even though it is possible to implement solutions to remind users in high risky time periods, attackers could still reverse engineer that approach and explore other time periods for carrying out attacks. Besides, it is not possible to keep users alert all the time. Hence, we consider the study of phishing email's distribution as having a medium opportunity for impact.\\

Due to the lack of attention and research on stage two factors, it is difficult to make a confident argument about whether these variables influence users' phishing susceptibility. We see this as a great opportunity for future research to investigate the situational phishing variables and other potential variables, such as the users' feelings and stress. Besides, there are other environmental variables such as distraction, noise, lighting, and temperature around the workplace that should also be studied, because these variables could contribute to influencing users' task performance \cite{hygge2001effects}.

\begin{tcolorbox}[breakable]
\textbf{Take Away 4:} There is a huge gap in Stage two, where we see many potential research opportunities that should be considered in the future.
\end{tcolorbox}


\subsection{Stage three: in-the-moment state}
The in-the-moment state refers to individuals' state when dealing with a specific potential phishing email: from the user side, it refers to how the users interact with the email; from the context side, it includes the design of the phishing emails and how the design may influence users. For a particular email checking session, the users' behaviour patterns are carried from Stage two, but the phishing email content may further bias their judgement towards trusting or distrusting the message, resulting in different behaviours. A well-constructed phishing email may utilize persuasion principles \cite{cialdini2009influence}, and manipulate the message and aesthetics to trick the users into performing actions the phishers desire. It is worth noting that with the implementation of these phishing techniques, the attackers may persuade the users to act emotionally and impulsively, leading to emotional judgement. Overall, the better the attackers can use these techniques effectively, the better they can conduct a successful attack. Concerning the users' response, how the users interact with the phishing email also accounts for their susceptibility. For instance, if users spend time and effort reading the message, they will perform better in phishing detection \cite{canfield2016quantifying}. As Table \ref{tab:factorCoverage} shows, both the email and user interaction variables are widely studied. However, understanding these variables is not sufficient to reduce phishing susceptibility. More research would be needed to study approaches to keep users alert when necessary.

\subsubsection{Cognitive effort}

When users read phishing emails, the amount of cognitive effort they spend on understanding the message directly contributes to their performance. In this context, the cognitive effort includes awareness, attention, and elaboration. 

Awareness describes the state where individuals are conscious about something. Parsons et al. \cite{parsons2013phishing} conducted an email management study that compared the performance of informed and not informed participants. The informed group was aware of the experiment's purpose (hence primed for phishing), and they performed significantly better than the not-informed group. This experiment demonstrated how awareness plays an essential role in phishing. Awareness sometimes can influence the amount of attention individuals spend on the email, which affects the amount of information they can perceive. With lower attention, individuals tend to focus on the visual cues that catch their eyes and make judgements based on their feelings and intuition (it is also referred to as peripheral information processing \cite{norris2019psychology}). Conversely, with greater attention, users would exhibit more analytical thinking and concentrate more on the message delivered, thus perform better in the detection \cite{canfield2016quantifying, molinaro2019using} (central information processing \cite{norris2019psychology}). As suggested by Canfield et al. \cite{canfield2016quantifying}, users need to be somewhat suspicious \textit{before} they start to treat the email as phishing. This has been referred to as the cognitive shift \cite{wash2020experts}, where users change their mindset from focusing on understanding the email to raised suspicion and investigating its legitimacy. Elaboration takes this one step further by consciously making connections between the cues and their knowledge and experience. Individuals who carry out a higher level of elaboration are less susceptible to phishing attacks as they are more likely to detect the threat \cite{musuva2019new, harrison2016individual}. The amount of cognitive effort users spend on a potential phishing email is an important predictor of the user's likelihood of identifying the attack. It is essential to understand why users change their effort in reading emails, so that we can develop interventions to motivate the users to pay attention when necessary. Hence, we label cognitive effort as having major opportunities for impact.

Vishwanath et al.'s simulation phishing experiment \cite{vishwanath2018suspicion} studied the cognitive process when checking emails using the Heuristic Systematic Model (HSM), which is a type of dual-process model. Vishwanath et al.'s study suggests that more heuristic processing leads to lower suspicion, whereas more systematic processing leads to higher suspicion. Since the activation of the systematic process requires cognitive effort \cite{sloman1996empirical}, an increase in cognitive effort would lead to more systematic reasoning of the email's legitimacy. Wang's study \cite{wang2017coping} on coping responses in phishing detection also supports this finding. Higher cognitive effort and attention can lead to adaptive coping, whereas lower cognitive effort would lead to maladaptive coping. With adaptive coping, users are more likely to engage in task-focused coping that actively seeks cues to determine emails legitimacy, thus having a higher chance of detecting phishing emails. In contrast, where maladaptive coping, or emotion-focused and avoidance coping, were maximized, it leads to effortless but biased judgements based on emotions and feelings.

\begin{tcolorbox}[breakable]
\textbf{Take Away 5:} Cognitive effort is an essential variable of phishing susceptibility. However, it is important to realize that we can not expect users to always pay attention to their emails because their primary goal is efficiently getting work done instead of checking for emails' legitimacy. Hence, the research on cognitive effort should also focus on studying how to help the users \textit{efficiently} identify emails' legitimacy without impacting their working efficiency.
\end{tcolorbox}

\subsubsection{Persuasion methods}
Attackers can adopt many different methods when constructing phishing emails. These persuasion methods aim to bias the user into performing quick and often emotional responses instead of logical processes that take time and effort. The implementation of proper persuasion principles can make phishing emails look trustworthy, the selection of email stories can raise the users' interest, and the arousal of the emotions can make the users' response emotional, leading to risky behaviour. We rate this variable with a medium opportunity for impact because the users have no control over the types of phishing emails they receive. However, it is valuable to notify the existence of certain persuasion methods used in the email so that the users are aware of the potential risks. Further, the study of the technical aspect of phishing could benefit from studying persuasion methods because the result could be used in machine learning or natural language processing to help develop better phishing filters and other countermeasures.

The psychological persuasion principles proposed by Cialdini \cite{cialdini2009influence} have been studied in the context of phishing emails in recent years. There are six principles: \textit{authority, consistency, liking, reciprocation, scarcity and social proof}. These principles were first studied in the phishing domain in 2014 by Wright et al. \cite{wright2014research} to analyze which principles are more effective in persuading users to click on the links embedded in phishing emails. Since then, several other studies have considered these principles in their phishing studies \cite{burda2020testing, lin2019susceptibility, taib2019social}. The effectiveness of these principles can differ in different contexts. Suppose a particular principle is extensively used in a short time frame (such as the \textit{authority} principle \cite{wright2014research}), the users would be more familiar and alert about the existence of such attacks. As a result, the community would build up resilience against this type of phishing attack and reduce phishing susceptibility. One study found that, between 2010 and 2015, the phishing email trend shows an increased the use of \textit{consistency} and \textit{scarcity}, and a decreased use of \textit{reciprocation} and \textit{social proof} \cite{zielinska2016temporal}. The effectiveness of the principles also depends on the email content. For instance, it is more reasonable to use the \textit{authority} principles in an email about security updates or password change than an email that promotes a product. If the principles are not properly used with appropriate content, they could backfire and make the users suspicious. Lawson et al.'s email management study \cite{lawson2020email} shows that the use of different persuasion principles can influence phishing susceptibility and individuals' judgment preference. When the \textit{authority} and \textit{scarcity} principles are used, users are more likely to classify the email as phishing. Conversely, when the \textit{liking} principle is used, users would tend to treat the email as legitimate. Also, younger users show greater susceptibility to \textit{scarcity} than older users, and older users show higher susceptibility to \textit{reciprocation} and \textit{liking} strategies than younger users \cite{lin2019susceptibility}. It is worth noting that other social engineering principles such as Gragg et al.'s principles \cite{gragg2003multi}, and Stajano et al.'s principles \cite{stajano2011understanding} have also been studied in relation to phishing, and have been merged and reviewed by Ferreira et al. \cite{ferreira2015principles}. These findings suggest that more research is needed to keep track of the effectiveness of these persuasion principles in phishing email construction.

Since different user groups are interested in various types of emails, the selection of correct \emph{target interest} is essential. If the users are not interested in the content, they may not read the phishing email even if the email is persuasively constructed. A theme that raises users' interests can result in a higher chance of deceiving the victim \cite{goel2017got, holm2014empirical}. House and Raja \cite{house2019phishing} discovered that when the email is important to the users, they will be more involved in reading the email; hence, they are more likely to be emotionally aroused and respond to the email. In other words, how well the attackers can tell a story that interests the target can influence how likely the target would fall for the attack. Studies have shown that a loss-based email (threatening the loss of properties/valuables) is generally more persuasive and seen as trustworthy than a reward-based email (gaining benefits) \cite{goel2017got, williams2019persuasive}. The degree of the loss also contributes to the persuasiveness of the email, where lower damage/loss can lead to a higher persuasiveness, resulting in a higher victimization \cite{luo2013investigating}. Interestingly, Harrison et al. \cite{harrison2016individual} found this to be insignificant and did not influence the attention to the phishing email, suggesting more study is necessary to explore these inconsistent findings.


As discussed above, the selection of email themes can influence users' decision making. This is partially because the content can cause the users to be emotionally aroused. As has been studied, individuals' task performance is associated with their emotions \cite{cai2011modeling}. Cai and lin \cite{cai2011modeling} conducted a driving simulation experiment and found an inverted U-shaped relationship between task performance and emotional arousal, and between task performance and emotional valence (negative, neutral, or positive). The result implies that optimal task performance would occur when both arousal and valence are neutral. Therefore, attackers may construct phishing emails that emotionally arouse the user to reduce their phishing detection performance. Emotions can be induced by the manipulation of the story (gaining goods or loss/protection of assets) and language of the message (positive or negative tone) \cite{tian2019effects}. Tian and Jensen's study \cite{tian2019effects} on positive and negative emotions (using loss/gain-based themes) found that emails that induce positive emotion are more effective in convincing users to click on the embedded links than negative emotions. This finding is in line with Forgas and East's theory \cite{forgas2008being} that a happy mood can make people more gullible than a neutral or sad mood. Also, Harrison et al. \cite{harrison2016individual} showed that individuals being aroused by different emotional stimuli can cause them to focus on different cues and information and interpret the message differently. Fear is related to the promotion of protective motivations \cite{jansen2018persuading}, and individuals with higher fear-arousal are less likely to respond to a phishing email and provide personal information \cite{house2019phishing}.

\subsubsection{Visual Presentation}
The design choices associated with the visual presentation of phishing emails have also been widely studied, and they do contribute to the success of a phishing attack. Pfeffel et al. \cite{pfeffel2019user} conducted an email management study that uses eye-tracking devices to investigate where users look when reading emails. They found that ordinary users spend most of their time in the body of the mail, whereas experts pay more attention to the header and attachment. When the users are focused on the main body, they can be distracted by the visual presentation. As has been found by Harrison et al. \cite{harrison2015examining}, phishing emails containing richer information (including logos, images, and aesthetic designs) are two times more likely to succeed than emails with lower richness (i.e., lack of images and logos). With richer visual presentation, users are more likely to rely on these visual cues (misleading cues) to heuristically determine the email's legitimacy, resulting in higher victimization \cite{harrison2015examining}. Similarly, emails with high authentic design cues (e.g., suggestive of legitimate organizations) have higher persuasiveness and are thus rated more trustworthy than emails with low authentic cues \cite{williams2019persuasive}. When phishing emails are not well crafted, grammar and spelling errors are common. However, it is still unclear whether even this would influence phishing susceptibility \cite{blythe2011f, vishwanath2011people, harrison2016individual, parsons2016users}. The visual presentation of the email can impact how individuals judge its trustworthiness. If the user feels the email is trustworthy, they tend to classify it as a legitimate email, and if the email is not trustworthy, the email will likely be classified as phishing \cite{downs2007behavioral}. Attackers would usually manipulate the email sender address and embedded URLs to look similar to the legitimate one \cite{pearson2017click, janet2008analysis}; if the users are not cautious enough, they could misjudge the email as legitimate. Similar to persuasion methods, we label visual presentation as a medium opportunity of impact because even though visual presentation can influence phishing success rate, it is not difficult for attackers to mimic legitimate email's visual presentation. Discussion of the details of secure email and assured provenance is beyond the scope of this paper, but assuring legitimacy remains challenging despite much work: see Clark et al. \cite{clarkemailsok}. We do see design opportunities to use heuristics to signal concern to users, perhaps changing visual presentation to draw attention. The adversarial and adaptable nature of phishing, however, makes success of such an approach uncertain.




\subsection{Summary}
To sum up, phishing susceptibility can be influenced by variables from three stages: the long-term stable stage, the situational stage, and the in-the-moment stage. Among these studies, it is most certain that knowledge and cognitive effort are negatively related to phishing susceptibility. Higher knowledge level and reading effort lead to higher phishing detection, thus reducing phishing susceptibility. Also, a well-constructed and highly-contextualized phishing email is more susceptible than a poorly constructed phishing email. However, we are still unsure whether demographic differences and personalities are factors that lead to a different performance in phishing detection. These uncertainties and unknowns suggest the study of phishing is not complete, and still more future research is needed to help users protect themselves from phishing attacks.

Our model can also adapt to other elements in the security protection chain, such as the existing countermeasures. As shown in Table \ref{tab:modelWithCountermeasures}, Stage one includes most of the technical countermeasures, such as blocklisting, phishing filters, and software infrastructure associated with the email communication system. On the user side, direct training provides the knowledge users need to identify phishing emails. For the third stage, the web filters add another layer of security when users try to open the links in emails. In contrast, embedded training will educate the users after they fall for a simulated phishing attack, to reduce the users' phishing susceptibility for subsequent attempts. From Table \ref{tab:modelWithCountermeasures}, it is obvious that there are no existing countermeasures that focus on the Stage two. It would be valuable to consider the situational characteristics when assessing the users' susceptibility against phishing at that moment, and investigate approaches to notify the users about unusual and suspicious emails in their inboxes \textit{before} they open them.

\begin{table}[ht]
\centering
\caption{PSM in relation to existing countermeasures}
\label{tab:modelWithCountermeasures}
\begin{tabular}{|p{0.8cm}|p{3cm}|p{1cm}|p{2cm}|}
\hline
 & Stage 1 & Stage 2 & Stage 3 \\ \hline
User & \begin{tabular}[t]{@{}l@{}}- Direct user training\\ - Security policies\end{tabular} &  & - Embedded user training \\ \hline
Context & \begin{tabular}[t]{@{}l@{}}- Infrastructure\\ - Blocklisting\\ - Phishing filters\\ - Two-factor authentication\end{tabular} &  & - Web filters \\ \hline
\end{tabular}
\end{table}

\section{Phishing Research Gaps} \label{gaps}
As we developed the model, 
In this section, we  discuss the research gaps that emerge from Table \ref{tab:factorCoverage}. We believe addressing these research gaps has high potential for helping users reduce their phishing susceptibility. It is clear from Table \ref{tab:factorCoverage} that most studies have focused on Stage one and Stage three variables. Even so, more research is needed to address the inconclusive findings in those studies. Furthermore, only one reviewed paper studied Stage two variables as their primary goal. This unbalanced distribution of research demonstrates the potential for future work to fill the gaps and understand phishing better. 

In recent years, technical advancements have made the detection of users' biological signals easy and convenient. This opens a new direction for future research on phishing susceptibility. We see a great opportunity for research to adopt this technology in understanding user behaviour and using psychological theories to reason about the findings. Regarding phishing detection, we found several areas that have been studied in psychology, and can be applied to human-centred phishing studies to help to explain the variations in users' phishing susceptibility. As such, we hope future studies can focus on users' mental processes to understand the variables that can influence their decision and impact their phishing detection performance. From the reviewed studies, we also found a lack of discussion on tools or systems that aim to help users determine email legitimacy when phishing emails arrived in their mailbox. This kind of system is necessary as we should not expect ordinary users to have the knowledge and effort to detect phishing emails, so we need such systems to help the users make correct decisions.


We provide a summary table of the variables with their opportunities for impact in the Appendix.


\subsection{Users' situational state}

Since users are the last line of defence, their situational state during email checking sessions is essential to their performance against phishing. For instance, users' phishing susceptibility is dependent on their perception, and response to the email. Therefore, the study of the variables that influence users' perception and behaviours in those moments is essential to understand why users fall for the attack. The idea that our mind is a dual process system has become popular in recent decades. Sloman \cite{sloman1996empirical} characterizes this as our mind having two reasoning systems. One is the associative system (also called the heuristic system), it is fast and effortless, and the reasoning is based on heuristics and intuitions. The other is the rule-based system (also called the analytic system), and it is slow and effortful, and will make decisions based on systematically processing the information. 
Related findings have been reported for some time, and were prominently discussed by Kahneman \cite{kahneman2011thinking}.
In relation to phishing, how users reason about the email perceived is influenced by their situational state, such as their mood, stress level, and mental and physical condition. In addition, their willingness and effort spent reading emails can impact the activation of the two reasoning systems. Below, we list several variables that could influence users' decision-making process.

\subsubsection{Emotions}

To date, most of the research on emotion in phishing has focused on emotional arousal induced by users' beliefs or the message delivered. However, email is not the only source of induced emotion; the users can also form emotions prior to or during the email checking task. Since emotions can influence task performance \cite{cai2011modeling}, it is reasonable to believe that the emotions that are carried over to the email checking session can influence the users' phishing detection performance. Moreover, when emotionally aroused users read a phishing email that further arouses their emotion, the resulting behaviour would be worth studying. For instance, it would be interesting to study how users in a happy mood would respond to phishing emails that intended to induce negative emotions and vice versa. The study of the carried over emotions can further build up our understanding of how emotions affect phishing susceptibility. We believe the understanding of users' emotions when checking email can help us predict users' susceptibility to phishing at that moment. Future research could consider adopting facial emotion recognition technology, or using biological signals to measure the users' emotions in real-time, and provide users with feedback when necessary regarding their situational phishing susceptibility, to provide an extra layer of awareness against phishing. Therefore, we rate the study of emotions as having many opportunities for impactful solutions.

\subsubsection{Stress and mental fatigue}
Similarly, stress and mental fatigue can influence phishing susceptibility by affecting the users' cognition and behaviour. Both stress and mental fatigue can lead to a reduction in any task performance \cite{van2017effects, lorist2000mental, boksem2005effects}. An increase in stress level results in a reduction in productivity \cite{halkos2010effect, imtiaz2009impact}. Stress can impair rational decision making by suppressing the activation of more systematic and controlled processing and motivate the use of heuristic and intuitive processing. It is worth noting that stress is not always bad for us. A certain amount of ``good stress" (or \textit{eustress}) can help users increase their adaptive capacity; it has been proven to improve productivity \cite{le2003eustress, kupriyanov2014eustress}.

We speculate that these variables would have a similar impact on phishing detection, with higher stress and mental fatigue leading to lower phishing detection performance. We believe these variables can be measured using portable sensory devices such as health trackers, eye trackers or even smartphones. Commercial health trackers (such as the Polar OH1 \cite{polar} and Empatica E4 \cite{empatica}) can monitor the users' heart rate (HR), photoplethysmogram (PPG) and galvanic skin response (GSR) signals, which can be used to calculate individuals' stress level or sleeping quality, and reflect their mental state. With these technologies in mind, it is now feasible to explore this new area of how mental states such as stress or mental fatigue can influence individuals' phishing susceptibility. Therefore, we consider studying stress and mental fatigue as having many opportunities for impact, as the design could potentially adapt to sensed stress and fatigue. 


\subsection{Distraction}
Another variable that could influence users' decision making is distraction. Due to the popularity of digital devices, people are overloaded with information. As a result, many users start multitasking to ``get things done efficiently". Since it is difficult for humans to focus on multiple tasks, multitasking is usually referred to as task switching \cite{adler2012juggling}. Studies have shown that task switching can reduce task performance and quality \cite{adler2012juggling, junco2012class}. Individuals could be switching between reading emails and other activities, and even replying to messages may constitute significant task-switching. However, there is a cost when individuals are switching between tasks \cite{monsell2003task}. Apart from the extra time required to get oneself re-familiar with a previous task, there is attention residual left in mind, which can disrupt the acquisition of information from the potential phishing email, thus influence decision making. Consequently, our mind needs to actively keep the unrelated information out to ensure a high concentration on the current task. By assessing how individuals check their emails (with or without task switching), we might observe a difference in phishing detection performance. 

Similar to task switching, mind-wandering is a concept of shifting attention from task-related processing to unrelated thoughts; it is a lapse of executive control \cite{mcvay2010does}. People mind-wander more when they are bored, stressed, dislike their current task, or are bad at their current task \cite{kane2007whom}. Email content may be associated with users' tendency to mind-wander. Research has shown that negative emotions can lead to a higher tendency to mind-wander and pay less attention to the current task \cite{smallwood2009shifting, taruffi2017effects}. This implies that if the users were in a bad mood before checking their emails, or the current email makes them uncomfortable, they may have a higher tendency to mind-wander. Since mind-wandering is associated with variables that lead to reducing task performance, it is reasonable to expect that a higher tendency of mind-wandering can reduce individuals' phishing detection performance and increasing phishing susceptibility.

We consider distraction as having medium opportunities of impact because the study of distractions can help identify potential distractors in the situation, leading to better management of distractions that help reducing phishing susceptibility. However, these distractions are sometimes unavoidable. Also, we acknowledge that the data collection process for these areas could be complex.

\subsection{Access methods}
As discussed earlier, users' behaviour could be influenced by the way they interact with their emails. Blythe's experiment \cite{blythe2011f} on blind people demonstrates a new study area. The way users perceive the message can account for different phishing susceptibility. Considering blind users, they are less susceptible to phishing attacks because most attackers do not design their phishing emails with blind users in mind. Therefore, the phishing cues that ordinary users do not easily detect can be obvious for blind users. The different interpretations of the information can make certain aspects of the message stand out more, thus affecting how the information is perceived. For typical users, they may not be able to interact with emails by audio as efficiently as blind users, but other alternative access methods might similarly allow identification of  phishing attacks. Research has found that information presentation can influence users' judgment and decision making \cite{kelton2010effects, engin2017information}. Since email clients can have different layouts and interface designs on distinct platforms (smartphone and computer), they could respond differently in terms of their information security awareness and behaviours. Future research could investigate the influence of varying the email clients' interface and platform and help the users focus on the correct cues in specific platforms that lead to identifying the email's legitimacy. As we mentioned, we consider this variable as having many opportunities for impact because understanding how the elements in different email clients and platforms influence user decision-making can lead to improving the user interface (UI) and user experience (UX) designs of the email system to reduce phishing susceptibility.

\subsection{Assistant tools}
As mentioned in Take Away 5, the users' primary goal for email is communication. Unfortunately, after emails reach their mailbox, it becomes  the users' responsibility to detect and deal with phishing emails. Most of the studies focusing on training users to detect phishing neglect that checking for email legitimacy is mostly a secondary task. Hence, more attention should be given to developing tools to help users reduce the effort in detecting phishing emails. There are existing studies that explore this problem. As pointed out by Brinton et al. \cite{brinton2016users}, warning messages of a general nature reduce their effectiveness after a few repetitions due to habituation. Brinton's solution is to create polymorphic warning messages, which can reduce the rate of forming habituation. Still, these polymorphic warning messages will lose their effectiveness eventually. Petelka et al. \cite{petelka2019put} conducted a study where they developed a technical intervention to force the users to wait a few seconds before they could click on embedded links. The result demonstrates that this is effective in reducing phishing susceptibility. But, with forced attention, the users have to spend a certain amount of time reading the URL (or simply waiting) before performing any further actions, thereby reducing their working efficiency and productivity. Also, this intervention can lead to negative emotions such as annoyance, and further motivate impulsive behaviours, resulting in misjudgment of the email's legitimacy. Nevertheless, such technical interventions show an opportunity and a need for more research on other approaches to help users assess their email for phishing. Similar to the access methods, developing tools would have many opportunities for impact because such work will add an extra layer of protection before users perform risky actions, thus reducing phishing susceptibility.\\

Most of the gaps we discovered can be summarized as either investigating how much information the users perceive (whether due to their cognitive effort or the technology used) or the variables that influence their mental conditions. These all come down to exploring ways to efficiently help users detect phishing cues, even when they are not in their best state. The research gaps discussed in this section may not be the full list, but we hope the list provides new insight and research directions for a better understanding of phishing.

\section{Conclusion} \label{conclusion}

Phishing is a growing cybersecurity issue that can target both industry and individuals. Although there are technical interventions that aim to reduce phishing susceptibility, these technologies can not fully prevent phishing emails from reaching end users. Therefore, understanding why and how users are susceptible to phishing attacks is essential. In the present paper, we proposed our PSM (Phishing Susceptibility Model) covering the life-cycle of phishing emails, and providing a foundation for identifying all phishing susceptibility variables. The PSM is specifically designed for studying phishing susceptibility.  Using this model, we have categorized variables contributing to phishing susceptibility, and identified several research gaps that deserve to be addressed. We also provided a radical practical impact analysis for assessing the value of research, and a quality of evidence analysis to guide future research and inform experiment quality. We recommend that the community should pay more attention to experiment design to ensure that research findings are produced with high quality, reliability, and generalizability. We consider this systematization to be a useful assessment of published research to date. We hope that by systematizing phishing variables according to our model, we provide inspiration and a basis for future research.


%







\bibliographystyle{IEEEtran}
\bibliography{references}

\section{Appendix}

In Table \ref{tab:opportunities}, we present a summary of the variables discussed in the paper with their opportunities for impact. 


\begin{table*}[ht]
\begin{adjustwidth}{-0.3cm}{}
\caption{Summary of the variables' opportunities for impact}
\label{tab:opportunities}
\begin{tabular}[t]{|p{1.3cm}|p{1.7cm}|p{1.6cm}|p{12.5cm}|}
\hline
 & \textbf{Variable} & \textbf{Opportunities of impact} & \textbf{Reasons} \\ \hline

\multirow[t]{4}{*}{\textbf{Stage one}}
 & Knowledge & Many & \begin{tabularx}{\hsize}{p{12cm}} \cmark consistent, confirmed finding with direct impact on phishing susceptibility\\ \cmark potential future direction of how to efficiently help users gain phishing related knowledge\\ \cmark could influence the in-the-moment state\end{tabularx} \\ \cline{2-4} 

 & Perception and beliefs & Medium & \begin{tabularx}{\hsize}{p{12cm}} \cmark research shows that beliefs do have an impact on phishing susceptibility\\ \xmark but the effect of some beliefs are still unclear (contradictory/insignificant findings)\\ \cmark could influence the in-the-moment state\end{tabularx} \\ \cline{2-4} 
 
 & Personality & Low & \begin{tabularx}{\hsize}{p{12cm}} \xmark contradictory/insignificant results\\ \xmark can be difficult to change a person's personality\\ \cmark can help identify susceptible user groups, develop specific training for corresponding populations\end{tabularx} \\ \cline{2-4} 
 
 & Habits & Medium & \begin{tabularx}{\hsize}{p{12cm}} \xmark can be difficult to change a person's habits\\ \cmark not all email reading habits are bad\\ \cmark potential future direction of studying approaches to help users build up good habits\end{tabularx} \\ \hline
 
\multirow[t]{4}{*}{\textbf{Stage two}}
 & \begin{tabular}{@{}l} Access\\methods \end{tabular} & Many & \begin{tabularx}{\hsize}{p{12cm}} \cmark for typical users, it refers to different approaches of gathering information\\ \cmark potential future direction of studying user behaviours when using different platforms for checking emails\\ \cmark can lead to the study of UI, UX design changes of the email system to improve performance\end{tabularx} \\ \cline{2-4} 

 & Email load & Medium & \begin{tabularx}{\hsize}{p{12cm}} \xmark current norms are instantaneous delivery whereas delays may be benefitial in some cases\\ \cmark potential future direction of studying the management of emails according to its source\end{tabularx} \\ \cline{2-4} 
 
 & Workplace management & Medium & \begin{tabularx}{\hsize}{p{12cm}} \cmark studies could look at ways to improve the workplace atmosphere to raise awareness etc.\\ \xmark involve team management level operations instead of individual level.\end{tabularx} \\ \cline{2-4} 
 
 & Distribution time & Medium & \begin{tabularx}{\hsize}{p{12cm}} \cmark can help identify risky time periods where phishing emails are likely to occur, and keeping users alert during the time periods\\ \xmark attackers control the distribution time, users can not be alerted all the time\end{tabularx} \\ \hline
 
\multirow[t]{3}{*}{\textbf{Stage three}}
& Cognitive effort & Many & \begin{tabularx}{\hsize}{p{12cm}} \cmark consistent, confirmed findings with direct impact on phishing susceptibility\\ \cmark potential future direction of developing interventions to motivate higher cognitive effort/attention when necessary\end{tabularx} \\ \cline{2-4} 

 & Persuasion methods & Medium & \begin{tabularx}{\hsize}{p{12cm}}  \cmark valuable in terms of reminding users of the existence of such persuasion methods to help users raise attention when necessary\\ \cmark valuable from the ML/NLP perspective in developing phishing countermeasures\end{tabularx} \\ \cline{2-4} 
 
 & \begin{tabular}{@{}l} Visual\\presentation \end{tabular} & Medium & \begin{tabularx}{\hsize}{p{12cm}}  \xmark too easy to replicate the visual presentation of a legitimate email\\ \cmark potential future direction of systematically manipulating the visual presentation based on legitimacy assessment\end{tabularx} \\ \hline
 
\multirow[t]{5}{*}{\textbf{\begin{tabular}{@{}l} Research\\Gaps \end{tabular}}}
& Emotions & Medium & \begin{tabularx}{\hsize}{p{12cm}} \cmark potentials of providing real-time feedback to help users when needed\\ \cmark emotion has been proven to influence phishing susceptibility, but part of this area is not studied (carry over emotion)\end{tabularx} \\ \cline{2-4} 

 & Stress / mental fatigue & Many & \begin{tabularx}{\hsize}{p{12cm}} \cmark potentials of providing real-time feedback to help users when needed\\ \cmark influence task performance, thus should have similar effects on phishing susceptibility\end{tabularx} \\ \cline{2-4} 
 
 & Distraction & Medium & \begin{tabularx}{\hsize}{p{12cm}} \cmark potentials of providing real-time feedback to help users when needed\\ \cmark influence task performance, thus should have similar effects on phishing susceptibility\\ \xmark could be difficult to measure\\ \xmark not controlled by the user\end{tabularx} \\ \cline{2-4} 
 
 & \begin{tabular}{@{}l} Access\\methods \end{tabular} & Many & \begin{tabularx}{\hsize}{p{12cm}}mentioned above\end{tabularx} \\ \cline{2-4}
 
 & Assistant tools & Many & \begin{tabularx}{\hsize}{p{12cm}} \cmark study of assistant tools when checking emails can provide an additional layer of protection\\ \cmark there are existing studies in this area, but still no effective solutions to the problem, hence require more future study\end{tabularx} \\ \hline
 
\end{tabular}
\end{adjustwidth}
\end{table*}

\end{document}